\documentclass[envcountsame]{llncs}

\usepackage{amsmath}
\usepackage{amssymb}
\usepackage{tikz}
\usepackage{todonotes}
\usepackage{algorithm}
\usepackage{algpseudocode}
\usepackage{booktabs}

\usetikzlibrary{shapes,positioning,calc,arrows,snakes}
\tikzstyle{vert}=[minimum size=0.6cm,inner sep=0cm,draw,circle]
\tikzstyle{vert2}=[vert, node distance=1.5cm]

\DeclareMathOperator{\play}{Play}
\DeclareMathOperator{\maxio}{MaxIo}

\newcommand{\veven}{V_{\textnormal{Even}}}
\newcommand{\vodd}{V_{\textnormal{Odd}}}
\newcommand{\seven}{\Sigma_{\textnormal{Even}}}
\newcommand{\sodd}{\Sigma_{\textnormal{Odd}}}
\newcommand{\weven}{W_{\textnormal{Even}}}
\newcommand{\wodd}{W_{\textnormal{Odd}}}

\newcommand{\vstart}{\mathsf{start}_{v}}
\newcommand{\vend}{\mathsf{end}_{v}}
\newcommand{\ustart}{\mathsf{start}_{u}}

\newcommand{\vals}{\textnormal{Vals}}
\newcommand{\sinit}{\sigma_{\textnormal{init}}}
\DeclareMathOperator{\pri}{pri}
\DeclareMathOperator{\val}{Val}

\DeclareMathOperator{\br}{br}
\DeclareMathOperator{\maxdiff}{Maxdiff}

\DeclareMathOperator{\suc}{succ}
\newcommand{\allse}{\textnormal{All}_\textnormal{Even}}
\newcommand{\allso}{\textnormal{All}_\textnormal{Odd}}

\newcommand{\nats}{\mathbb N}
\newcommand{\ints}{\mathbb Z}

\begin{document}

\title{Efficient Parallel Strategy Improvement for Parity Games\thanks{This
work was supported by EPSRC grant EP/P020909/1 ``Solving Parity Games in Theory
and Practice.''}}
\author{John Fearnley}
\institute{Department of Computer Science, University of Liverpool, UK}

\maketitle

\begin{abstract}
We study strategy improvement algorithms for solving parity games. While these
algorithms are known to solve parity games using a very small number of
iterations, experimental studies have found that a high step complexity causes
them to perform poorly in practice. In this paper we seek to address this
situation. Every iteration of the algorithm must compute a best response, and
while the standard way of doing this uses the Bellman-Ford algorithm, we give
experimental results that show that one-player strategy improvement
significantly outperforms this technique in practice. We then study the best way
to implement one-player strategy improvement, and we develop an efficient
parallel algorithm for carrying out this task, by reducing the problem to
computing prefix sums on a linked list. We report experimental results for these
algorithms, and we find that a GPU implementation of this algorithm shows a
significant speedup over single-core and multi-core CPU implementations.

\end{abstract}

\section{Introduction}

\paragraph{\bf Parity games.}

A parity game is a zero-sum game played on a finite graph between two players
called Even and Odd. Each vertex of the graph is labelled with an integer
\emph{priority}. The players move a token around the graph to form an infinite
path, and the winner is determined by the \emph{parity} of the largest priority
that is visited infinitely often: Even wins if and only if it is even.

Parity games have attracted much attention in the verification community,
because they capture the expressive power of nested least and greatest fixpoint
operators, as formalized in the modal $\mu$-calculus and other fixpoint
logics~\cite{GTW02}. In particular, deciding the winner in parity games is
polynomial-time equivalent to checking non-emptiness of non-deterministic parity
tree automata, and to the modal $\mu$-calculus model checking, two fundamental
algorithmic problems in automata theory, logic, and
verification~\cite{EJS93,Sti95,GTW02}.

\paragraph{\bf Strategy Improvement.}

We study \emph{strategy improvement} for solving parity
games, which is a local search technique that iteratively
improves the strategy of one of the two players until an optimal strategy is
found. Much like the simplex method for linear programming, and policy iteration
algorithms for MDPs, strategy improvement algorithms can solve
large parity games in a very small number of iterations in practice. The first
strategy improvement algorithm devised specifically for parity games was given
by V\"oge and Jurdzi\'nski~\cite{VJ00}, and since then several further
algorithms have been proposed~\cite{BV07,L08,schewe08}.

Every strategy improvement algorithm uses a \emph{switching rule} to decide how
to proceed in each step. Theoretically, the best known switching rule is the
\emph{random-facet} rule, which provides a $2^{O(\sqrt{n \log n})}$ upper bound
on the number of strategy improvement iterations~\cite{MSW96}. However, this is
a \emph{single switch} rule, which which only switches one edge in each
iteration. In practice, we would expect an arbitrarily chosen initial strategy
to differ from an optimal strategy by $O(n)$ edges, and so a single switch rule
will necessarily cause the strategy improvement algorithm to take at least
$O(n)$ iterations.

In this paper, we focus on the \emph{greedy all-switches} switching rule, which
switches every vertex that can be switched in each iteration. This rule has been
found to perform very well in practice, and as our experimental results confirm,
greedy all-switches strategy improvement can solve games with more than ten
million vertices in under one-hundred iterations.

\paragraph{\bf Practical aspects of strategy improvement.}

Although strategy improvement can solve large games using only a handful of
iterations, experimental work has found that it performs very
poorly in practice. For example, Friedmann and Lange performed an experimental
study~\cite{FL09} in which the all-switches variant of the V\"oge-Jurdzi\'nski
algorithm was compared with Jurdzi\'nski's
\emph{small-progress measures} algorithm~\cite{J00} and Zielonka's
\emph{recursive} algorithm~\cite{Z98}. They found that, in some cases, the
V\"oge-Jurdzi\'nski algorithm takes longer than an hour to solve games with
under one-hundred thousand vertices, whereas the recursive algorithm can scale
to problems that are an order of magnitude larger.

The reason for this is that, although the algorithm uses a very small number of
iterations, the cost of performing each step is very high. In particular, the
V\"oge-Jurdzi\'nski algorithm, which has served as the standard benchmark for
strategy improvement algorithms, has a step complexity of $O(n^2)$, even in
games with a small number of priorities\footnote{This is because the algorithm
requires that every vertex has a distinct priority, and so comparing two
valuations requires $O(n)$ time.}.

In fact, there are existing algorithms that avoid this high step complexity.
Bj{\"{o}}rklund, Sandberg and Vorobyov present an algorithm whose step
complexity is $O(n \cdot d)$, where $d$ is the number of distinct priorities
used in the game. While $d$ can be as large as $n$ in the case where every
vertex has a distinct priority, in practice $d$ is often a very small constant
such as $2$ or $4$. Luttenberger observed~\cite{L08} that a particularly simple
algorithm is obtained if one combines the Bj\"orklund-Vorobyov strategy
improvement algorithm for mean-payoff games~\cite{BV07}, with the discrete
valuation used by the V\"oge-Jurdzi\'nski algorithm. This algorithm also has
$O(n \cdot d)$ step complexity, and is the one that we will focus on in this
paper.

\paragraph{\bf Our contribution.}

Our goal in this paper is to show that all-switches strategy improvement can be
used in practice to solve large parity games efficiently. As we have mentioned
above, the number of iterations needed by the algorithm is usually tiny, and so
our effort is dedicated towards improving the cost of computing each step.
The main contributions of this paper are:


\paragraph{Best response computation.}
In each iteration of strategy improvement, the algorithm has a strategy for one
of the two players, and must compute a best response strategy for the opponent.
This can be a very expensive operation in practice. For the algorithm studied in
this paper, this boils down to solving a solving a shortest paths problem that
can contain negative weights. The natural approach is to apply the Bellman-Ford
algorithm. However, the first contribution of this paper is to show that there
is a better approach: best responses can be computed using a one-player version
of strategy improvement.


The performance  of strategy improvement algorithms on shortest paths problems
was studied by Cochet-Terrasson and Gaubert~\cite{CG00}. While they showed that
the number of improvement iterations is at most $O(|V||E|)$, their experimental
results on random graphs found that strategy improvement was outperformed by the
Bellman-Ford algorithm. They found that, while they typically both take the same
number of iterations, one iteration of strategy improvement is more expensive
than one iteration of Bellman-Ford. 

Nevertheless, for the case of parity games, we give experimental evidence to
show that one-player strategy improvement outperforms the Bellman-Ford algorithm
when computing best responses. The experimental data
shows that part, but not all, of this improvement is due to the fact that we can
initialize the algorithm with the best response from the previous iteration. 


\paragraph{A parallel algorithm for strategy improvement.}
Once we fix the decision to use one-player strategy improvement to compute best
responses, we turn our attention towards the best way to implement this. In recent years, hardware manufacturers have made little progress
in speeding up single-core CPU workloads, but progress continues to be made by
adding more cores to CPUs. Moreover, GPUs continue to be made more powerful, and
the rise of general purpose computing on GPUs has found many prominent
applications, for example, in the training of deep neural networks. For this
reason, we argue that good parallel implementations are required if we are to
use an algorithm in practice.

The second contribution of this paper is to develop an efficient parallel
algorithm for computing a strategy improvement iteration. The decision to use
one-player strategy improvement to compute best responses means that the only
non-trivial task is to compute the \emph{valuation} of a pair of strategies. We show
that this task can be reduced to an instance of \emph{list ranking}, a
well-studied problem that requires us to compute the prefix-sum of a linked
list. The first work optimal parallel algorithm for list ranking was given by
Cole and Vishkin~\cite{CV89}. However, their algorithm is complex and difficult
to implement in practice. Helman and J\'aj\'a give a simpler randomized
algorithm that is work efficient with high probability~\cite{HJ99}, and in
particular it has been shown to work well on modern GPU hardware~\cite{WJ12}. We
give a modification of the Helman-J\'aj\'a algorithm that can be used to compute
a valuation in a parity game.

\paragraph{Experimental results.} 
We have produced CPU and GPU implementations of the aforementioned parallel
algorithm. The third contribution of this paper is to provide experimental
results. We use the recently developed benchmark suite
of Keiren~\cite{K15}, which unlike previous benchmarks from
PGSolver~\cite{FL09}, contains large parity games derived from real verification
tasks. 

We find that our implementation scales to parity games with
tens of millions of vertices, and that the limiting factor is memory rather than
run time. We also compare a single-threaded sequential CPU implementation with a
multi-threaded parallel CPU implementation and a GPU implementation, which both
use list ranking algorithm described above. While the parallel CPU
implementation fails to deliver a meaningful speedup, the GPU implementation
delivers an average speedup of 10.37.

\paragraph{\bf Related work.}

Strategy improvement originated from the \emph{policy
iteration} algorithms that are used to solve Markov decision
processes~\cite{puterman94}, and can be seen as a generalisation of this method
to the two-player setting. The method was first proposed by Hoffman and Karp in
order to solve two-player concurrent stochastic games~\cite{HK66}. It was then
adapted by Condon~\cite{C90} to solve simple-stochastic games, and by Puri to
solve discounted games~\cite{puri95}. Parity games can be reduced in polynomial
time to discounted and simple-stochastic games~\cite{J98,ZP96}, so both of these
algorithms could, in principle, be used to solve parity games, but both
reductions require the use of large rational numbers, which makes doing so
impractical.

The greedy all-switches switching rule has received much attention in the past.
Its good experimental performance inspired research
into whether it always terminates after polynomially many
iterations. However, Friedmann showed that this was not the case~\cite{F11}, by
giving an example upon which the algorithm takes exponential time.
Recently, it has even been shown that deciding whether a given strategy is
visited by the algorithm is actually a PSPACE-complete problem~\cite{FS16}.

There has been much previous work on solving parity games in parallel. Most of
the work so far has focused on the small progress measures algorithm~\cite{J00},
because it can be ,implemented in parallel in an straightforward way. In the
first paper on this topic, van de Pol and Weber presented a multi-core
implementation of the algorithm~\cite{PW08}, and Huth, Huan-Pu Kuo, and Piterman
presented further optimizations to that algorithm~\cite{HKP11}. Two papers have
reported on implementations on the parallel Cell processor used by the
Playstation 3~\cite{K09,B10}.

For parallel implementations of strategy improvement, there are two relevant
papers. Hoffman and Luttenberger have given GPU implementations of various
algorithms for solving parity games~\cite{HL13}. In particular, they implemented
the strategy improvement algorithm that is studied in this paper, but they used the
Bellman-Ford algorithm to compute best responses. Meyer and Luttenberger have 
reported on a GPU implementation of the Bj\"orklund-Vorobyov strategy
improvement algorithm for mean-payoff games~\cite{ML16}.







\section{Preliminaries}

\paragraph{\bf Parity games.}

A parity game is played between two players called Even and Odd. Formally, it is
a tuple $\mathcal{G} = (V, \veven, \vodd, E, \pri)$, where $(V, E)$
is a directed graph. The sets $\veven$ and $\vodd$ partition $V$ into the
vertices belonging to player Even, and the vertices belonging to player Odd,
respectively. The \emph{priority} function $\pri : V \rightarrow \nats$ assigns
a positive natural number to each vertex. We define $D_\mathcal{G} = \{p \in
\nats \; : \; \pri(v) = p \text{ for some } v \in V\}$ to be the set of
priorities that are used in $\mathcal{G}$. We make the standard assumption that
there are no terminal vertices, which means that every vertex is required to
have at least one outgoing edge. 

A \emph{positional strategy} for player Even is a function that picks one
outgoing edge for each Even vertex. More formally, a positional strategy for Even is a
function $\sigma : \veven \rightarrow V$ such that, for each $v \in \veven$ we
have that $(v, \sigma(v)) \in E$. Positional strategies for player Odd are
defined analogously.  We use $\seven$ and $\sodd$ to denote the set of
positional strategies for players Even and Odd, respectively. Every strategy
that we consider in this paper will be positional, so from now on, we shall
refer to positional strategies as strategies.

A \emph{play} of the game is an infinite path through the game. More precisely,
a play is a sequence $v_0, v_1, \dots $ such that for all $i\in \nats$ we have
$v_i \in V$ and $(v_i, v_{i+1}) \in E$. Given a pair of strategies $\sigma \in
\seven$ and $\tau \in \sodd$, and a starting vertex $v_0$, there is a unique
play that occurs when the game starts at $v_0$ and both players follow their
respective strategies. So, we define $\play(v_0, \sigma, \tau) = v_0, v_1,
\dots$, where for each $i \in \nats$ we have $v_{i+1} = \sigma(v_i)$ if $v_i \in
\veven$, and $v_{i+1} = \tau(v_i)$ if $v_i \in \vodd$.

Given a play $\pi = v_0, v_1, \dots$ we define:
\begin{equation*}
\maxio(\pi) = \max \{ p \; : \; \exists \text{ infinitely many } i \in \nats \text{ s.t. }
\pri(v_i) = p\},
\end{equation*}
to be the largest priority that occurs \emph{infinitely often} along $\pi$. 
The winner is determined by the parity of this priority:
a play $\pi$ is \emph{winning} for player Even if $\maxio(\pi)$ is
even, and we say that $\pi$ is winning for Odd if $\maxio(\pi)$ is odd.

A strategy $\sigma \in \seven$ is a \emph{winning strategy} for a vertex $v \in
V$ if, for every (not necessarily positional) strategy $\tau \in \sodd$, we have that $\play(v, \sigma,
\tau)$ is winning for player Even. Likewise, a strategy $\tau \in \sodd$ is a
winning strategy for $v$ if, for every (not necessarily positional) strategy $\sigma \in \seven$, we have
that $\play(v, \sigma, \tau)$ is winning for player Odd. The following
fundamental theorem states that parity games are \emph{positionally determined}.
\begin{theorem}[\cite{ej91,mostowski91}]
The set of vertices~$V$ can be partitioned into
\emph{winning sets} $(\weven, \wodd)$, where Even has a positional
winning strategy for all $v \in \weven$, and Odd has a positional winning
strategy for all $v \in \wodd$.
\end{theorem}

\noindent The computational problem that we are interested in is, given a parity
game, to determine the partition $(\weven, \wodd)$.

\section{Strategy Improvement}

In this section, we describe the strategy improvement algorithm that we will
consider in this paper. The algorithm, originally studied by
Luttenberger~\cite{L08}, is a combination of the Bj\"orklund-Vorobyov strategy
improvement algorithm for mean-payoff games~\cite{BV07}, with the discrete
strategy improvement valuation of V\"oge and Jurdzi\'nski~\cite{VJ00}. Strategy
improvement algorithms select one of the two players to be the strategy
improver. In this description, and throughout the rest of the paper, we will
select player Even to take this role.

\paragraph{\bf A modified game.}

At the start of the algorithm, we modify the game by introducing a new
\emph{sink} vertex $s$ into the graph. For each vertex $v$ of the Even player,
we add a new edge from $v$ to the sink. The idea is that, at any point player
Even can choose to take the edge to $s$ and terminate the game. The owner and
priority of $s$ are irrelevant, since the game stops once $s$ is reached. 

\paragraph{\bf Admissible strategies.} 

A strategy $\sigma \in \seven$ is said to be \emph{admissible} if player Odd
cannot force and odd cycle when playing against $\sigma$. More formally,
$\sigma$ is admissible if, for every strategy $\tau \in \sodd$ we have that
$\play(v, \sigma, \tau)$ either arrives at the sink $s$, or that
$\maxio(\play(v, \sigma, \tau))$ is even. The strategy improvement algorithm
will only consider admissible strategies for player Even. 

\paragraph{\bf Valuations.}
The core of a strategy improvement algorithm is a \emph{valuation}, which
measures how good a given pair of strategies is from a given starting vertex.
For our algorithm, the valuation will count how many times each priority occurs
on a given path, so formally a valuation will be a function of the form
$D_\mathcal{G} \rightarrow \ints$, and we define $\vals_\mathcal{G}$ to be the
set of all functions of this form.

Given an admissible strategy $\sigma \in \seven$ for Even, a strategy $\tau
\in \sodd$ for Odd, and a
vertex $v \in V$, we define the \emph{valuation function} $\val^{\sigma,
\tau}(v) : V \rightarrow \vals_\mathcal{G} \cup \{\top\}$ as follows.
\begin{itemize}
\item If $\pi = \play(v, \sigma, \tau)$ is infinite, then we define
$\val^{\sigma, \tau}(v) = \top$

\item If $\pi = \play(v, \sigma, \tau)$ is finite, then it must end at the sink
$s$.  The valuation of $v$ will count the number of times that each priority
appears along $\pi$. Formally, if $\pi = v_0, v_1, \dots, v_k, s$, then for
each $p \in D_V$ we define a valuation $L \in \vals_\mathcal{G}$ as follows:
\begin{equation*}
L(p) = \bigl| \{ i \in \nats \; : \; \pri(v_i) = p \} \bigr|.
\end{equation*}
We set $\val^{\sigma, \tau}(v) = L$.
\end{itemize}
Observe that, since $\sigma$ is an admissible strategy, $\val^{\sigma, \tau}(v)
= \top$ implies that $\play(v, \sigma, \tau)$ is winning for Even.

Next, we introduce the operator $\sqsubseteq$ which will be used to compare
valuations. We define $L \sqsubseteq \top$ for every $L \in
\vals_\mathcal{G}$. When we compare two valuations, however, the procedure is more
involved. Let $L_1, L_2 \in \vals_g$ be two valuations. If $L_1 = L_2$ then $L_1
\sqsubseteq L_2$ and $L_2 \sqsubseteq L_1$. Otherwise, we define
$\maxdiff(L_1, L_2)$ to be the largest priority $p$ such that $L_1(p) \ne
L_2(p)$. Then, we have that $L_1 \sqsubseteq L_2$ if and only if one of the
following is true:
either $p = \maxdiff(L_1, L_2)$ is even and $L_1(p) < L_2(p)$,
or $p = \maxdiff(L_1, L_2)$ is odd and $L_1(p) > L_2(p)$.

\paragraph{\bf Best responses.}
Given an admissible strategy $\sigma \in \seven$, a \emph{best response} is a
strategy $\tau \in \sodd$ that minimizes the valuation of each vertex. More
formally, we define, $\br(\sigma) \in \sodd$ to be a strategy with the property
that $\val^{\sigma, \br(\sigma)}(v) \sqsubseteq \val^{\sigma, \tau}(v)$ for every
strategy $\tau \in \sodd$ and every vertex $v$. If there is more than one such
strategy, then we pick one arbitrarily. 
Although it is not immediately clear, it can be shown that there is a single
strategy $\tau \in \sodd$ that simultaneously minimises the valuation of all
vertices. 
Strategy improvement only ever considers
an admissible strategy $\sigma$ played against its best response, so we define
the shorthand $\val^{\sigma} = \val^{\sigma, \br(\sigma)}$.


\paragraph{\bf The algorithm.}
We are now ready to describe the strategy improvement algorithm. It begins by
selecting the following \emph{initial strategy} for Even. We define $\sinit \in
\seven$ so that $\sinit(v) = s$ for all $v \in \veven$. Note that there is no
guarantee that $\sinit$ is admissible, because there may be a cycle with odd
parity that contains only vertices belonging to player Odd. So, a preprocessing
step must be performed to eliminate this possibility. One simple preprocessing
procedure is to determine the set of vertices from which Odd can avoid visiting
an Even vertex, and to insert enough dummy Even vertices into this subgame to
prevent Odd from forming a cycle. As it happens, none of the games considered in
our experimental study require preprocessing, so this is not a major issue in
practice.

In each iteration, strategy improvement has a strategy for the improver. The
first step is to compute the set of \emph{switchable} edges for this strategy.
An edge $(v, u)$ is switchable in strategy $\sigma$ if $u \ne \sigma(v)$ and
$\val^{\sigma}(\sigma(v)) \sqsubset \val^{\sigma}(u)$. We define
$\mathcal{S}^{\sigma}$ to be the set of edges that are switchable in $\sigma$. 

The algorithm selects a non-empty subset of the switchable edges and
\emph{switches} them. We say that a set of edges $S \subseteq E$ is a
\emph{switchable set} if, for every pair of edges $(v, u), (v', u') \in S$, we
have has $v \ne v'$, that is, $S$ does not contain two outgoing edges for a
single vertex. If $S$ is a switchable set and $\sigma$ is a strategy, then we
can \emph{switch} $S$ in $\sigma$ to create the new strategy $\sigma[S]$ where,
for every vertex~$v$:
\begin{equation*}
\sigma[S](v) = \begin{cases}
u & \text{$(v, u) \in S$,} \\
\sigma(v) & \text{otherwise.}
\end{cases}
\end{equation*}

The key property of strategy improvement is that, if $S \subseteq \mathcal{S}^{\sigma}$ is
a switchable set that contains only switchable edges, then we have that
$\sigma[S]$ is better than $\sigma$ in the $\sqsubseteq$ ordering. Formally, this means that $\val^{\sigma}(v)
\sqsubseteq \val^{\sigma[S]}(v)$ for all vertices $v$, and there exists at least
one vertex for which we have $\val^{\sigma}(v) \sqsubset \val^{\sigma[S]}(v)$.

The strict improvement property mentioned above implies that the
algorithm cannot visit the same strategy twice, so it must eventually terminate.
The algorithm can only terminate once it has reached a strategy with no
switchable edges. We can use this strategy to determine winning sets for both
players. That is, if $\sigma^*$ is a strategy with no switchable edges, then we
can prove that:
$\weven = \{ v \in V \; : \; \val^{\sigma^*}(v) = \top \}$, and
$\wodd = \{ v \in V \; : \; \val^{\sigma^*}(v) \ne \top \}$.


Luttenberger has given a direct proof that the algorithm is correct~\cite{L08}.
Actually, a simple proof of correctness can be obtained directly from the
correctness of Bj\"orklund-Vorobyov (BV) algorithm. It is not difficult
to show that if we turn the parity game into a mean-payoff game using the
standard reduction, and then apply the BV algorithm to the resulting mean-payoff
game, then the BV algorithm and this algorithm will pass through exactly the
same sequence of strategies. 

\begin{theorem}
\label{thm:algmain}
The following statements are true.
\begin{itemize}
\item For every strategy $\sigma \in \seven$ there is at least one best response
$\tau \in \sodd$.
\item Let $\sigma$ be a strategy, and let $S \subseteq \mathcal{S}^{\sigma}$ be
a switchable set that contains only switchable edges.
We have $\val^{\sigma}(v)
\sqsubseteq \val^{\sigma[S]}(v)$ for all vertices $v$, and there exists at least
one vertex for which we have $\val^{\sigma}(v) \sqsubset \val^{\sigma[S]}(v)$.
\item Let $\sigma$ be a strategy that has no switchable edges.
We have
$\weven = \{ v \in V \; : \; \val^{\sigma}(v) = \top \}$, and
$\wodd = \{ v \in V \; : \; \val^{\sigma}(v) \ne \top \}$.
\end{itemize}
\end{theorem}

\paragraph{\bf Switching rules.} 

Strategy improvement always switches a subset of switchable edges, but we have
not discussed \emph{which} set should be chosen. This decision is delegated to a
\emph{switching rule}, which for each strategy picks a subset of the switchable
edges. In
this paper we will focus on the \emph{greedy all-switches} rule, which  always
switches every vertex that has a switchable edge. If a vertex has more than one
switchable edge, then it picks an edge $(v, u)$ that maximizes
$\val^{\sigma}(u)$ under the $\sqsubseteq$ ordering (arbitrarily if there is
more than one such edge.)

\section{Computing Best Responses}

To implement strategy improvement, we need a method for computing best
responses. Since we only consider admissible strategies for Even, we know that
Odd cannot create a cycle with odd parity, and so computing a best response
simply requires us to find a shortest-path from each vertex to the sink, where
path lengths are compared using the $\sqsubseteq$ ordering. Any vertex that has
no path to the sink is winning for Even. The obvious way to
do this is to apply a shortest-paths algorithm. Note that odd priorities
correspond to negative edges weights, so a general algorithm,
such as the Bellman-Ford algorithm, must be applied.

\paragraph{\bf One-player strategy improvement.}

In this paper, we propose an alternative: we will use one-player strategy
improvement equipped with the greedy-all switches rule. We say that an edge $(v,
u)$ is \emph{Odd-switchable} if $v \in \vodd$ and $\val^{\sigma,
\tau}(\sigma(v)) \sqsupset \val^{\sigma, \tau}(u)$. To find a best response
against a fixed admissible strategy $\sigma \in \seven$, the algorithm starts
with an arbitrary Odd strategy $\tau \in \sodd$, and repeatedly switches
Odd-switchable edges until it arrives at an  Odd strategy in which there are no
Odd-switchable edges. 

It is not difficult to see that if $\tau$ has no Odd-switchable edges when
played against $\sigma$, then it is a best response against $\sigma$,
because a strategy with no Odd-switchable edges satisfies the Bellman optimality
equations for shortest paths. 


One-player strategy improvement algorithms for solving shortest paths problems
were studied by Cochet-Terrasson and Gaubert~\cite{CG00}. In particular, they
proved that the all-switches variant of the algorithm always terminates after at
most $O(|V||E|)$ steps. Hence, we have the following lemma.

\begin{lemma}
\label{lem:brsi}
Let $\sigma$ be an admissible strategy. One-player strategy
improvement will find a best-response against $\sigma$ after at most
$O(|V||E|)$ iterations.
\end{lemma}

\paragraph{\bf The algorithm.}

We can now formally state the algorithm that we will study. Given a 
strategy $\sigma \in \seven$, let $\allse(\sigma)$ be
the function that implements the greedy all-switches switching rule as described
earlier. Moreover, given a pair of strategies $\sigma \in
\seven$ and $\tau \in \sodd$, let $\allso(\sigma, \tau)$ be a set $S$ of
Odd-switchable edges $(v, u)$ such that there is no edge $(v, w) \in E$ with
$\val^{\sigma, \tau}(u) \sqsupset \val^{\sigma, \tau}(w)$, and such that each
vertex has at most one outgoing edge in~$S$.

\begin{algorithm}[h]
\caption{The strategy improvement algorithm}
\label{alg:si}
\begin{algorithmic}
\State Initialize $\sigma := \sinit$ and set $\tau$ to be an arbitrary strategy.
\Repeat
\Repeat 
    \State Compute $\val^{\sigma, \tau}(v)$ for every vertex $v$.
    \State Set $\tau := \tau[S_\text{Odd}]$ where $S_\text{Odd} = \allso(\sigma, \tau)$.
\Until {$S_\text{Odd} = \emptyset$}
    \State Set $\sigma := \sigma[S_\text{Even}]$ where $S_\text{Even} = \allse(\sigma)$.
\Until {$S_\text{Even} = \emptyset$}
\end{algorithmic}
\end{algorithm}

The inner loop computes
best responses using one-player strategy improvement, while the outer loop
performs the two-player strategy improvement algorithm. Note, in particular,
that after switching edges in $\sigma$, the first Odd strategy considered by the
inner loop is the best response to the previous strategy. 




\section{Parallel Computation of Valuations}

Most operations used by strategy improvement can naturally be carried out in
parallel. In particular, if we have already computed a valuation, then deciding
whether an edge is switchable at a particular vertex $v$, and finding the
switchable edge that has the highest valuation at $v$, are both local properties
that only depend on the outgoing edges of $v$. So these operations can trivially
be carried out in parallel. This leaves the task of
computing a valuation as the only task that does not have an obvious parallel
algorithm.

In this section, we give an efficient parallel algorithm for computing a
valuation. Given two strategies $\sigma \in \seven$ and $\tau \in \sodd$ in a
game $\mathcal{G}$, we show how computing $\val^{\sigma, \tau}(v)$ can be
parallelized in a work efficient manner. There is an obvious sequential
algorithm for this task that runs in time $O(|V| \cdot |D_\mathcal{G}|)$ which
works backwards on the tree defined by $\sigma$ and $\tau$ and counts how many
times each priority appears on each path to $s$. Every vertex not found by this
procedure must have valuation $\top$.






\paragraph{\bf List ranking.}
The idea of our algorithm is to convert the problem of computing a valuation,
into the well-known problem of computing prefix-sums on a linked list, which is
known as \emph{list ranking}. We will then adapt the efficient parallel
algorithms that have been developed for this problem.

Given a sequence of integers
$x_0, x_1, x_2, \dots, x_k$, 
and a binary associative operator $\oplus$,
the \emph{prefix-sum} problem requires us to compute a
sequence of integers $y_0, y_1, y_2, \dots, y_k$ such that $y_i = x_1 \oplus x_2
\oplus \dots \oplus x_{i-1}$. If the input sequence is given as an array, then
efficient parallel algorithms have long been known~\cite{LF80}.

If the input sequence is presented as a linked-list, then the problem is called
the list ranking problem, and is more challenging. The first work optimal
parallel algorithm for list ranking was given by Cole and Vishkin~\cite{CV89}.
However, their algorithm is complex and difficult to implement in practice.
Helman and J\'aj\'a give a simpler randomized algorithm that is work efficient
with high probability~\cite{HJ99}.

\begin{theorem}[\cite{HJ99}]
There is a randomized algorithm for list ranking that, with high probability,
runs in time $O(n/p)$ whenever $n > p^2 \ln n$, where $n$ denotes the
length of the list, and $p$ denotes the number of processors.
\end{theorem}

We now give a brief overview of the algorithm, as we will later modify it
slightly. A full and detailed description can be found in~\cite{HJ99}. The
algorithm works randomly choosing $s = \frac{n}{p \log n}$ elements of the list
to be \emph{splitters}. Intuitively, each splitter defines a sublist that begins
at the splitter, and ends at the next splitter that is encountered in the list
(or the end of the list). These sublists are divided among the processors, and
are ranked using the standard sequential algorithm. Once this has been
completed, we can create a \emph{reduced} list, in which each element is a
splitter, and the value of each element is the prefix-sum of the corresponding
sublist. The reduced list is ranked by a single processor, again using the
standard sequential algorithm. Finally, we can complete the list ranking task as
follows: if an element $e$ of the list has rank $x_r$ in its sublist, and the
splitter at the start of sublist has rank $x_s$ in the reduced list, then the
rank of $e$ is $x_s \oplus x_r$.

\paragraph{\bf Pseudoforests and Euler tours.}
We now show how the problem of computing a valuation can be reduced to list
ranking. Let $\mathcal{G}^{\sigma, \tau} = (V, \veven, \vodd, E^{\sigma,
\tau}, \pri)$ be the game $\mathcal{G}$ in which every edge not used by $\sigma$ and
$\tau$ is deleted. Since each vertex has exactly one outgoing edge in this
game, the partition of $V$ into $\veven$ and $\vodd$ are irrelevant, and we
shall treat $\mathcal{G}^{\sigma, \tau}$ has a graph labelled by priorities.

First, we observe that $\mathcal{G}^{\sigma, \tau}$ is a
\emph{directed pseudoforest}.
The set of vertices whose valuation is not $\top$ form a
directed tree rooted at $s$. For these vertices, our task is to count the number
of times each priority occurs on each path to the sink, and hence compute a
valuation.
Each other vertex is part of a \emph{directed pseudotree}, which is a
directed tree in which the root also has exactly one outgoing edge that leads
back into the tree. 
Since we deal only with admissible strategies, every vertex in a pseudotree has
valuation $\top$.

\begin{figure}
\begin{center}
\begin{tikzpicture}
\node[vert] (s) {$a$};
\node[vert, below left of=s] (1) {$b$};
\node[vert, below right of=s] (2) {$c$};
\node[vert, below left of=1] (3) {$d$};
\node[vert, below right of=1] (4) {$e$};

\path[arrows={-stealth}]
    (1) edge (s)
    (2) edge (s)
    (3) edge (1)
    (4) edge (1)
    ;

\node[vert2, right of=s, node distance=5cm] (s2) {$a,b$};
\node[vert2, below left= 0.5cm and 0.5cm  of s2] (10) {$b, d$};
\node[vert2, below left= 0.5cm and -1.0cm of 10] (11) {$d, b$};
\node[vert2, right = 0.5cm of 10] (12) {$b, e$};
\node[vert2, below right = 0.5cm and 0.5cm of 12] (13) {$e, b$};
\node[vert2, below right = 0.5cm and 1cm of s2] (14) {$b, a$};
\node[vert2, below right = 0.5cm and 0.5cm of 14] (15) {$a, c$};
\node[vert2, right = 1.5cm of s2] (16) {$c, a$};

\path[arrows={-stealth}]
    (s2) edge (10)
    (10) edge (11)
    (11) edge (12)
    (12) edge (13)
    (13) edge (14)
    (14) edge (15)
    (15) edge (16)
    ;



\end{tikzpicture}
\end{center}
\caption{Converting a tree into a linked list using the Euler tour technique.
Left: the original tree. Right: the
corresponding linked list.}
\label{fig:euler}
\end{figure}
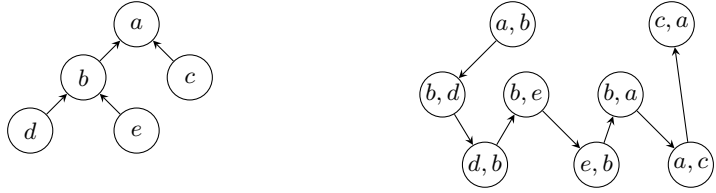

A standard technique for reducing problems on trees to list ranking is the
\emph{Euler tour} technique. We will describe this technique for the tree rooted
at $s$, and show that it can be used to compute a valuation. We will also use
the same technique for the other pseudo-trees in the graph, but since this
portion of the algorithm is not standard, we defer the description until later.

In order to compute a valuation for every vertex $v$ in the tree rooted at $s$,
we need to count the number of times that a given priority $p$ occurs on the
path from $v$ to the root. We create a linked list as follows. First we replace
each directed edge $(v, u)$ with two edges $(v, u)$ and $(u, v)$.
Then we select an Euler tour of this modified graph that starts and
ends at the root. We use this tour to create a linked list, in which each element of the
list an edge of the original tree, and the successors of each element are
determined by the Euler tour. The value associated with each element $e$ is
defined as follows: 
\begin{itemize}
\item If $e = (u, v)$, then the value of $e$ is $1$ if $\pri(v) = p$, and $0$
otherwise.
\item If $e = (v, u)$, then the value of $e$ is $-1$ if $\pri(v) = p$ and $0$
otherwise.
\end{itemize}
If we then compute a list ranking on this list using $+$ as the operator
$\oplus$, then the ranking of $(v, u)$ gives the number of times $p$ appears on
the path from $v$ to the sink. Obviously, to compute a valuation we must do the
above procedure in parallel for each priority in the game. 

\paragraph{\bf Formal reduction to list ranking.}

We now give a formal definition of the technique that we just described. 
Recall that $E^{\sigma, \tau}$ gives the edges chosen by $\sigma$ and $\tau$. We
define
\begin{equation*}
\overleftarrow{E}^{\sigma,\tau} = \{ (u, v) \; : \; (v, u) \in E^{\sigma,
\tau}\},
\end{equation*}
to be the set of reversed edges. We call each edge in $E^{\sigma, \tau}$ an
\emph{up} edge, since it moves towards the root, and correspondingly we call
each edge in $\overleftarrow{E}^{\sigma,\tau}$ a \emph{down} edge.
The set of elements in our linked list will be 
$L = E^{\sigma, \tau} \cup \overleftarrow{E}^{\sigma,\tau}$.

Next we define the successor function $\suc: L \rightarrow L \cup
\{\epsilon\}$, which gives the structure of the list, and where $\epsilon$ is
used to denote the end of the list. To do this, we take an arbitrary Euler tour
of the tree, and define $\succ$ to be the function that follows this tour.
Figure~\ref{fig:euler} gives an example of this construction.


In our overview, we described how to use list ranking to compute the number of
times a given priority $p$ appears on the path to the sink. In our formal
definition, we will in fact compute a full valuation with a single call to a
list ranking algorithm. To achieve this,
we define the weight function $w: L \rightarrow \vals_\mathcal{G}$ as follows. 
For each priority $p \in D_\mathcal{G}$, we first define two valuations $A_p,
A_{-p} \in \vals_\mathcal{G}$ so that, for
every $q \in D_\mathcal{G}$:
\begin{align*}
A_p(q) = \begin{cases}
1 & \text{if $q = p$,}\\
0 & \text{otherwise.}
\end{cases} &&
A_{-p}(q) = \begin{cases}
-1 & \text{if $q = p$,}\\
0 & \text{otherwise.}
\end{cases} 
\end{align*}
Then, for every list element $e = (u, v) \in L$:
if $e$ is an up edge then we set $w(e) = A_{-\pri(u)}$, and
if $e$ is a down edge then we set $w(e) = A_{\pri(v)}$.
Moreover, we define the binary operator $\oplus$ as follows. Given two
valuations $A_1, A_2$, we define $A_1 \oplus A_2 = A_3$ where for every priority
$p \in D_\mathcal{G}$ we have $A_3(p) = A_1(p) + A_2(p)$.

\paragraph{\bf Modifications to the Helman-J\'aj\'a algorithm.} 
We must also handle the vertices that lie in pseudotrees. Our
reduction turns every pseudotree into a pair of cycles. The Helman-J\'aj\'a
algorithm can be adapted to deal with these, by ensuring that if a cycle is
found in the reduced list, then all vertices on it are given a valuation of $\top$.
Moreover, some vertices may not be part of a reduced list, because they may be
part of a small pseudotree, and none of random splitters were in that
pseudotree. Since the Helman-J\'aj\'a always picks the head of the list to be a
splitter (in our case this would be an edge leaving the sink at the start of the
Euler tour), every vertex in the tree rooted at $s$ is in the reduced list. So
any vertex not part of a reduced list can be assigned valuation~$\top$.

\paragraph{\bf Constructing the list ranking instance in parallel.} 
Since at least one of $\sigma$ and $\tau$ will change between every iteration,
we must construct a new list ranking instance in every iteration of our
algorithm. Thus, in order to have a true parallel algorithm, we must be able to
carry out the reduction in parallel as well. 

We start by describing a sequential algorithm for the task. Each vertex in the
tree maintains two pointers $\vstart$ and $\vend$. Initially, $\vstart$
points to the down edge of $v$, and $\vend$ points to the up edge of $v$. Then,
in an arbitrary order, we process each vertex $v$, and do the following:
\begin{enumerate}
\item Determine the parent of $v$ in the tree, and call it $u$.
\item Connect the list element pointed to by $\ustart$ to the element pointed to
by $\vstart$.
\item Set $\ustart = \vend$.
\end{enumerate}
Once this has been completed, we then join the list element pointed to by $\vstart$
to the list element pointed to by $\vend$, for all vertices $v$.

Intuitively, this algorithm builds the tour of each subtree incrementally.
The second step adds the tour of the subtree starting at $v$ to the linked
list associated with $u$. The third step ensures that any further children of
$u$ will place their tours after the tour of the subtree of $v$.

For example, let us consider the tree and corresponding Euler tour given in
Figure~\ref{fig:euler}, and let us focus on the vertex $b$. Initially, $\vstart$
points to $(a, b)$, while $\vend$ points to $(b, a)$, which are the down and up
edges of $b$, respectively. Let us suppose that $d$ is processed before $e$.
When $d$ is processed, $(a, b)$ is connected to $(b, d)$ and $\vstart$ is
updated to point to $(d, b)$. Subsequently, when $e$ is processed $(d, b)$ is
connected to $(b, e)$, and $\vstart$ is updated to point to $(e, b)$.
Then, in the final step of the algorithm $(b, d)$ is connected to $(d, b)$ and
$(e, b)$ is connected to $(b, a)$. So, the linked list corresponding to the
subtree of $b$ (shown on the right in
Figure~\ref{fig:euler}) is created. Note that if $e$ was processed before $d$,
then a different linked list would be created, which would correspond to a
different Euler tour of the tree. From the point of view of the
algorithm, it is not relevant which Euler tour is used to construct the linked
list.

In theory, this algorithm can be carried out in parallel in $O(n/p)$ time and
$O(np)$ space by having each processor maintain its own copy of the pointers
$\vstart$ and $\vend$, and then after the algorithm has been completed, merging
the $p$ different sublists that were created.

In practice, the space blow up can be avoided by using atomic exchange 
operations, which are available on both CPU and GPU platforms. More precisely,
we can use an atomic exchange instruction to set $\ustart = \vend$, while
copying the previous value of $\ustart$ to a temporory variable, and then
connect the list element that was pointed to by $\ustart$ to $\vstart$.

\section{Experimental Results}

\paragraph{\bf Experimental setup.}

Our experimental study uses four implementations.
\begin{itemize}
\item GPU-LR: a GPU implementation that uses the list-ranking algorithm to
compute valuations. The GPU is responsible for ranking the sublists, while
ranking the reduced list is carried sequentially on the CPU.
\item CPU-Seq: a single-threaded implementation that uses the natural sequential
algorithm for computing valuations. 
\item CPU-LR: a multi-threaded CPU implementation that uses the list-ranking
algorithm to compute valuations. The sublists are ranked in parallel, while the
reduced lists is ranked by a single thread. 
\item Bellman-Ford: a single-threaded CPU implementation that uses the
Bellman-Ford algorithm to compute best responses.
\end{itemize}
All implementations are in C++, and the GPU portions are implemented using
NVIDIA CUDA. The code is publicly
available\footnote{\texttt{https://github.com/jfearnley/parallel-si}}. We also compare our results to PGSolver's
recursive algorithm, with all of PGSolver's heuristics disabled in order to
deliver a fair comparison. We chose the recursive algorithm because it was found
to be the most competitive in the previous experimental study of Friedmann and
Lange~\cite{FL09}.

\begin{table}
\resizebox{\textwidth}{!}{
\begin{tabular}{ll||rrrr||r}
Game & Property & Vertices & $\seven$ & $\sodd$ & Edges & Pris
\\
\midrule
CABP/Par 2 & branching-bisim & 167k & 79k & 88k & 434k & 2 \\ 
CABP/Par 1 & weak-bisim & 147k & 122k & 25k & 501k & 2 \\ 
ABP(BW)/CABP & weak-bisim & 157k & 129k & 27k & 523k & 2 \\ 
Elevator & fairness & 862k & 503k & 359k & 1.4m & 3 \\ 
Election & eventually-stable & 2.3m & 343k & 2.0m & 7.9m & 4 \\ 
Lift (Incorrect) & liveness & 2.0m & 999k & 999k & 9.8m & 4 \\ 
SWP/SWP 1 & strong-bisim & 3.8m & 1.5m & 2.2m & 11.5m & 2 \\ 
SWP  & io-read-write & 6.8m & 4.2m & 2.6m & 15.8m & 3 \\ 
CABP & io-receive & 7.0m & 5.2m & 1.8m & 24.9m & 2 \\ 
ABP/Onebit & weak-bisim & 8.3m & 7.2m & 1.1m & 31.3m & 2 \\ 
Hesselink/Hesselink & weak-bisim & 29.9m & 22.9m & 7.0m & 78.8m & 2 \\ 
SWP/SWP 2 & branching-bisim & 37.6m & 20.1m & 17.6m & 120.8m & 2 \\ 
ABP(BW)/Onebit & weak-bisim & 35.4m & 30.6m & 4.8m & 134.9m & 2 \\ 
SWP/SWP 3 & weak-bisim & 32.9m & 29.0m & 3.9m & 167.5m & 2\\
\end{tabular}
}
\vskip 0.5\baselineskip
\caption{The games that we consider in our experimental study. The table
displays the number of vertices, player Even vertices, player Odd vertices,
edges, and distinct priorities.}
\label{tbl:games}
\end{table}

For our benchmark games we utilise the suite that was recently developed by
Keiren~\cite{K15}. This provides a wide array of parity games that have been
used throughout the literature for model-checking and equivalence checking.
Since there are over 1000 games, we have selected a subset of those games to use
here, and these are shown in Table~\ref{tbl:games}.  In particular, we have
chosen a set of games that span a variety of sizes, and that cover a variety of
tasks from verification. We found that strategy improvement solves many of the
games in the suite in a very small number of iterations, so the results that we
present here focus on the games upon which strategy improvement takes the
largest number of iterations. The vast majority of the games in the suite have
between 2 and 4 priorities, and the ones that do not are artificially
constructed (eg. random games), so we believe that our sample is representative
of real world verification tasks.

The test machine has an Intel Core i7-4770K CPU, clocked at 3.50GHz (3.90GHz
boost), with 4 physical cores, and 16GB of RAM. The GPU is an NVIDIA GeForce
GTX 780, which has 2304 CUDA cores clocked at 1.05GHz and 3GB of RAM. At the
time of purchase in 2013, the CPU cost \pounds 248.20 and the GPU cost \pounds
444.94. Since the CPU has 8 logical cores with hyper-threading enabled, we use 8
threads in our CPU multi-threaded implementations. When benchmarking for time,
we ran each instance three times, and the reported results are the average of
the three. We implemented a time limit of 10 minutes. We only report the amount
of time needed to solve the game, discarding the time taken to parse the game.

\paragraph{\bf Best response algorithms.}

\begin{table}
\resizebox{\textwidth}{!}{
\begin{tabular}{l||rr||rr|rr|rr}
&  & Maj. & \multicolumn{2}{|c}{SI} & \multicolumn{2}{|c|}{SI-Reset} & \multicolumn{2}{c}{Bellman-Ford}\\
Game & Edges & Iter & Time (s) & Iter & Time (s) & Iter & Time (s) & Iter  \\
\midrule
CABP/Par 2 & 434k & 8 & 0.33 & 53 & 0.49 & 100 & 1.65 & 161 \\ 
CABP/Par 1 & 501k & 12 & 0.22 & 47 & 0.36 & 93 & 2.41 & 235 \\ 
ABP(BW)/CABP & 523k & 9 & 0.15 & 28 & 0.29 & 65 & 1.38 & 128 \\ 
Elevator & 1.4m & 33 & 13.18 & 231 & 17.88 & 364 & 216.36 & 2238 \\ 
Election & 7.9m & 77 & 41.43 & 364 & 57.66 & 585 & 157.5 & 842 \\ 
Lift (Incorrect) & 9.8m & 16 & 9.09 & 69 & 22.82 & 215 & 42.47 & 242 \\ 
SWP/SWP 1 & 11.5m & 8 & 14.69 & 58 & 22.25 & 93 & 71.21 & 152 \\ 
SWP  & 15.8m & 11 & 25.44 & 82 & 31.69 & 104 & 109.31 & 148 \\ 
CABP & 24.9m & 11 & 5.45 & 11 & 5.54 & 11 & 59.37 & 108 \\ 
ABP/Onebit & 31.3m & 20 & 34.15 & 57 & 93.78 & 234 & 494.97 & 604\\
\end{tabular}
}
\vskip 0.5\baselineskip
\caption{Experimental results comparing the algorithm used to compute a best
response. The algorithms are (1) SI: one-player strategy improvement (2) SI
(Reset): one-player strategy improvement starting from an arbitrary strategy (3)
Bellman-Ford. For each algorithm we report the total time and the total number
of iterations used by the best response algorithm.}
\label{tbl:br}
\end{table}

Our first experiment is to determine which method for computing best responses
is faster in practice. In this experiment we compare the single-core sequential
implementation of one-player strategy improvement (SI) against a single-core
sequential implementation of the Bellman-Ford algorithm. 

As we have mentioned, our one-player strategy improvement starts with the
optimal strategy against the previous strategy of the improver. To quantify the
benefit of this, we have also include results for a version of the one-player
strategy improvement algorithm that, at the start of each best response
computation, resets to the initial arbitrarily chosen strategy. We refer to this
as SI-Reset.

The results are displayed in Table~\ref{tbl:br}. We only
report results for games that Bellman-Ford solved within the 10 minute time
limit. We report the total number of \emph{major iterations}, which are the
iterations in which the improver's strategy is switched. The number of major
iterations does not depend on the algorithm used to compute best responses. For
each algorithm we report the overall time and the total number of iterations
used computing best responses.

Before discussing the results in detail we should first note that these results
paint a very positive picture for strategy improvement. All games were solved in
at most 77 major iterations, with most being solved with significantly fewer
major iterations. The number of iterations used on the Election instance was the
most that we saw over any instance in our study, including those that we do not
report here. This clearly shows that strategy improvement can scale to very
large instances.

Moving on to the choice of best response algorithm, the most striking feature is
that Bellman-Ford is on average 8.43 times slower than one-player strategy
improvement (min 3.80, max 16.42). Some of this difference can be explained by
the fact that Bellman-Ford is on average 1.72 times slower per iteration than
one-player strategy improvement (min 1.11, max 2.38), which may be due to
implementation inefficiencies. But most of the difference is due to the fact
that Bellman-Ford uses on average 5.30 times more iterations than one-player
strategy improvement (min 1.80, max 10.60).

The results with SI-Reset show that only some of this difference can be
attributed to reusing the previous best response as SI-Reset uses on average
2.05 times more iterations than SI (min 1.00, max 4.11).
Overall we found that SI
used an average of 5.49 iterations to compute each best response (min 1.0 max
9.9), which again indicates that this method can scale to very large games.

\paragraph{\bf Parallel implementations.}

\begin{table}
\resizebox{\textwidth}{!}{
\begin{tabular}{l||r||rr||rrrr}
&  & \multicolumn{2}{c||}{Iterations} & \multicolumn{4}{|c}{Time (s)} \\
Game & Edges  & Maj. & Tot. & GPU-LR & CPU-Seq & CPU-LR &
PGSolver \\
\midrule
CABP/Par 2 & 434.0k & 8 & 53 & 0.05 & 0.33 & 0.48 & 0.48 \\ 
CABP/Par 1 & 501.0k & 12 & 47 & 0.04 & 0.22 & 0.4 & 0.46 \\ 
ABP(BW)/CABP & 523.0k & 9 & 28 & 0.03 & 0.15 & 0.25 & 0.49 \\ 
Elevator & 1.4m & 33 & 231 & 0.87 & 13.18 & 11.56 & 13.23 \\ 
Election & 7.9m & 77 & 364 & 4.37 & 41.43 & 58.43 & 30.45 \\ 
Lift (Incorrect) & 9.8m & 16 & 69 & 0.79 & 9.09 & 9.35 & 40.76 \\ 
SWP/SWP 1 & 11.5m & 8 & 58 & 1.06 & 14.69 & 14.57 & 28.83 \\ 
SWP  & 15.8m & 11 & 82 & 2.71 & 25.44 & 35.01 & 201.83 \\ 
CABP & 24.9m & 11 & 11 & 0.39 & 5.45 & 5.33 & 134.33 \\ 
ABP/Onebit & 31.3m & 20 & 57 & 2.28 & 34.15 & 32.46 & ---$^\dagger$\\
Hesselink/Hesselink & 78.8m & 28 & 142 & ---$^\dagger$ & 318.98 & 299.43 & ---$^\dagger$ \\ 
SWP/SWP 2 & 120.8m & 10 & 99 & ---$^\dagger$ & 282.17 & 265.62 & ---$^\dagger$ \\ 
ABP(BW)/Onebit & 134.9m & 20 & 57 & ---$^\dagger$ & 147.03 & 142.35 & ---$^\dagger$ \\ 
SWP/SWP 3 & 167.5m & 10 & 71 & ---$^\dagger$ & 142.96 & 168.45 & ---$^\dagger$\\
\end{tabular}
}
\vskip 0.5\baselineskip
\caption{Experimental results comparing the running time of (1) GPU-LR: list
ranking on the GPU (2) CPU-Seq: a sequential CPU implementation (3) CPU-LR: list
ranking on a 4-core CPU (4) PGSolver: the recursive algorithm from PGSolver. $\dagger$ indicates a failure due to lack of memory.} 
\label{tbl:main}
\end{table}

Our second set of experimental results concerns our parallel implementation of
strategy improvement when best responses are computed by one-player strategy
improvement. The results are displayed in Table~\ref{tbl:main}. 

The first thing to note is that the parallel algorithm does not deliver good
performance when implemented on a CPU. On average the multi-threaded CPU list
ranking algorithm was 1.25 times \emph{slower} than the single-threaded sequential
algorithm (min 0.88, max 1.77). This can be partially explained by the fact that
the total amount of work done by the parallel algorithm is at least twice the
amount of work performed by the sequential algorithm, since turning the strategy
into a linked list doubles the number of vertices. 

On the other hand, the GPU implementation delivers a significant speedup.
To give a fair comparison between the GPU implementation and the CPU
implementations, we compute the ratio between the time taken by GPU-LR, and the
minimum of the times taken by CPU-Seq and CPU-LR. Using this metric we find that
the average speedup is 10.37 (min 5.54, max 14.21). The average speedup
increases to 12.17 if we discard instances with fewer than 1 million edges,
where setup overhead makes the GPU algorithm less competitive.

The downside to the GPU implementation is that games with more than about 32
million edges are too large to fit within the 3GB of memory on our test GPU.
Obviously, there is a cost trade off between the extra speed delivered by a GPU
and the cost of purchasing a GPU with enough memory. At the time of writing,
relatively cheap consumer graphics cards can be bought with up to 8GB of memory,
while expensive dedicated compute cards are available with up to 24GB of memory. 

Finally, we compare our results to PGSolver's recursive algorithm. Here, to have
a fair comparison, we should compare with the sequential CPU algorithm, as both
algorithms are single-threaded. Unfortunately PGSolver ran out of memory for the
very large games in our test set, but for the smaller games it can be seen that
CPU-Seq is always competitive, and in many cases significantly faster than
PGSolver.

\bibliographystyle{abbrv}
\bibliography{references}

\appendix

\end{document}